\documentclass[prb,showpacs]{revtex4}
\usepackage{bm} 
\usepackage{amsmath}
\usepackage{amssymb}
\usepackage{amsfonts}
\usepackage{amsmath}
\usepackage{graphicx}
\usepackage[backref,pdffitwindow,colorlinks,citecolor={red},linkcolor={blue}]{hyperref}

\begin{document}
\title{Tailored Optical Polarization in Nano-Structured Metamaterials}
\author{Bernardo  S. Mendoza}
\affiliation{Department of Photonics, Centro de Investigaciones en \'Optica,
  Le\'on, Guanajuato, M\'exico}
\author{W. Luis Moch\'an}
\affiliation{Instituto de Ciencias F{\'\i}sicas, Universidad Nacional 
  Aut\'onoma de M\'exico, Apartado Postal 48-3, 62251 Cuernavaca,
  Morelos, M\'exico.}

\begin{abstract}
A very efficient method for the calculation of the effective 
optical response of nano-structured composite systems allows the
design of metamaterials tailored for specific optical polarization
properties. We use our method to design 2D periodic arrays of
sub-wavelength dielectric
inclusions within nanometric supported metallic thin films which
behave as either an
almost perfect linearly dichroic system, as a controllable source of circular
polarized light, as a system with a large circular dichroism, or as a
circular polarizer. All of these systems may be tuned over a wide
energy range.
\end{abstract}
\maketitle 
\section{Introduction}
The calculation of the macroscopic electromagnetic response of binary
composite materials made up of inclusions of an ordinary material
within another has been explored since the nineteenth
century.\cite{Etopim(1977),Etopim(1993),Etopim(2002)} 
Techniques such as electron beam lithography have allowed the
fabrication of nano-structured systems with inclusions of specific
shapes.\cite{Noda(2003),Grigorenko(2005)} 
Similarly, ion milling techniques have produced high quality
periodic patterns of holes of various shapes
forming two-dimensional (2D) arrays.\cite{kleinPRL04,gordonPRL04}  
Therefore,  it is possible
to conceive and fabricate devices with novel and exotic macroscopic optical
properties.\cite{Pendry(2000)}  
For example, a negative refractive index has
been predicted and observed\cite{Shalaev(2005)} for a periodic 
metamaterial consisting of a dielectric matrix with a periodic lattice
of noble metal
inclusions of trapezoidal shape.\cite{Kildishev(2006)} Devices based
on other metamaterials have been proposed to manipulate  the
direction of 
propagation of electromagnetic waves and bend
their trajectories and to 
focus light in sub-wavelength regions using flat
lenses,\cite{Pendry(2000)} produce electromagnetic
cloaking\cite{Leonhardt(2006),Pendry(2006),Milton(2006)} and 
shielding.\cite{Feng(2008)} Furthermore, metamaterials built with conductors may
display hyperbolic dispersion relations\cite{Liu(2007)} which yield
singular densities 
of states. They also display plasmonic resonances which may be used to
guide electromagnetic energy\cite{Barnes(2003)} in directions that
may be controlled through the polarization of light.\cite{Lin(2013)}
Chiral plasmonic 
metamaterials\cite{Cui(2014)} have been proposed to detect circularly polarized
light.\cite{Li(2015)}
Thus, the
development of fast computational procedures for efficiently
obtaining the electromagnetic 
properties of new nano-structured systems has become very important.

In Ref. \onlinecite{Mendoza(2012)} we employed a scheme based on Haydock's
recursive method\cite{Haydock(1980)} and developed in
Refs. \onlinecite{mochanOE10} and \onlinecite{cortesPSS10} to obtain
within the long-wavelength approximation the optical properties of
systems with arbitrary geometry and composition. Among other applications,
we studied a
film made of a square lattice of dielectric elliptical cylinders and rectangular
prisms within a conducting matrix. We obtained a strong birefringent
and dichroic 
response, such as a range of frequencies for which rotating by 90$^\circ$
the angle of polarization of the incoming light could change the film
from  from being an almost
perfect reflector to being an almost perfect absorber. The frequency where this
behavior was displayed was easily tuned by geometrical modifications
such as rotating the base of the prisms or the axes of the
ellipses. 

The
tremendous speed improvement over other equivalent approaches such as that of
Ref. \onlinecite{ortizPRB09} allows calculations for 2D\cite{cortesPSS10} and
3D\cite{mochanOE10} structures of arbitrary geometry, including
interpenetrated inclusions, and allowing for dispersive and dissipative
components. That approach is based on a local field effect theory
\cite{mochanPRB85a} which incorporates into the macroscopic response
the spatial fluctuations of the microscopic electric field due to the
texture of the composite. Similar homogenization procedures are also
found in
Refs. \onlinecite{ortizPRB09,Perez(2006),krokhinPRB02,haleviPRL99,dattaPRB93}. 
Our calculations proceeded from digitized
images of the system, such as  a photograph or a 
drawing, which can be manipulated by standard software to explore the
influence of geometry on response. This allowed us to
obtain artificial materials with the sought optical
properties. As long as we consider only sub-wavelength lengthscales,
we cannot explore effects such as geometrically induced chirality
\cite{Cui(2014),zhangPRL09,xiongPRB10, Kaschke(2015)} or magnetism in
left-handed 
metamaterials, although we have developed a generalization
\cite{Perez-Huerta(2013)} of our recursive procedure
\cite{mochanOE10,cortesPSS10} and we have shown that a 
macroscopic approach can deal with lengthscales comparable to
wavelength and yield, for example, the photonic band structure of
the system. 

In this paper we explore the polarization acquired by light at
anisotropic thin films of
metamaterials, where we use this term to describe
nano-structured composites whose properties differ from those of the natural
materials of which it is manufactured. In a periodic composite we
could expect two origins for an anisotropic behavior, even when the
component materials are isotropic. One is due to
the periodic lattice and another due to the shape of the individual
inclusions. Consider an inclusion with a shape that has well
defined symmetry axes. If these coincide with the crystalline axes,
they would also coincide with the principal axes of the
macroscopic response. If however, the inclusions are rotated with respect to
the crystalline lattice,\cite{Wang(2015)} the principal axes of the macroscopic
response would also {\em rotate} to new directions that in general
will depend on the 
composition of the metamaterial and the frequency of the light, i.e., they
will not be geometrically defined. Something similar would occur if
the inclusions lack symmetry axes.
Furthermore, if one or both of the
components of the metamaterial are dissipative, then the principal
directions are given by complex vectors and the corresponding
normal modes would be in general elliptically polarized. In this paper we
explore the polarization acquired by light as it is reflected by or
transmitted through a thin film of a metamaterial made up of a
dissipative metal with dispersionless dielectric inclusions with a
simple shape that is not necessarily aligned with the crystalline
axes.

We obtained nano-structured metamaterial films which display extreme
linear dichroism and that mimic quarter wave plates that allow the
conversion of linear to circular polarization, as well as optimized
structures 
that yield a large circular dichroism and that produce circular
polarized light when illuminated with unpolarized light, and these
properties may be tuned over a wide tunable energy range.
 
The article is organized as follows. In Sec. \ref{theory} we
briefly present the 
theoretical approach used for the calculation of the macroscopic
dielectric response of the metamaterial, to describe the elliptical
polarization of the fields and normal modes and to calculate the
optical properties of the metamaterial. 
In Sec. \ref{res} we 
present results for two-dimensional
periodic structures with inclusions of different shapes and
orientations. Finally,
in Sec. \ref{conc}, we present our conclusions.   

\section{Theory}\label{theory}

\subsection{Macroscopic response}\label{mr}

\begin{figure}
  \includegraphics[width=.7\textwidth]{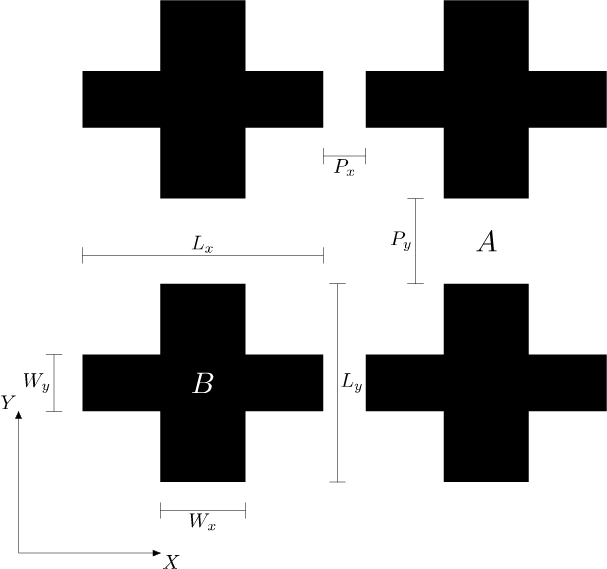}
  \caption{\label{cross} Geometry of a metamaterial made up of cross
    shaped holes in 
    a supported conducting film. We indicate the $X-Y$ axes, the
    length ($L_x$, $L_y$)
    of the beams, their width ($W_y$, $W_x$) and the widths of the
    conducting paths ($P_x$, $P_y$) between neighbor conducting
    rectangles. The crosses may additionally be rotated by an angle
    $\phi$ (not shown).}
\end{figure}

In this subsection we review the main theoretical steps in order to
calculate the macroscopic dielectric tensor, $\epsilon^{ij}_{M}$
following Refs.~\onlinecite{mochanOE10} and \onlinecite{cortesPSS10}.
We consider inclusions ($B$) embedded within a 
homogeneous material ($A$) forming a 2D periodic lattice in the $X-Y$
plane as shown, for example, in Fig.~\ref{cross}. The inclusions are
taken to be generalized  
{\em cylinders} with an arbitrary cross section and translationally
invariant along the $Z$ axis.  We  assume that each 
region $A$ or $B$ has a well defined 
dielectric function $\epsilon_\gamma$, where $\gamma=a,b$, which we
assume local and 
isotropic, and that the cross section of the inclusions and the
lattice periodicity 
are much smaller the free wavelength of light $\lambda_0=2\pi c/\omega$
with $c$ the speed of light in vacuum and $\omega$ the frequency. For
visible and near-infrared light, this implies that the inclusions must
be of nanometric size.

The {\em microscopic} response is then described by
\begin{equation}\label{er}
  \epsilon(\mathbf{r}) = \epsilon_a - B(\mathbf{r}) \epsilon_{ab}
\end{equation}
where $\epsilon_{ab}\equiv\epsilon_a-\epsilon_b$ and
$B(\mathbf{r})$ 
is the periodic characteristic function for the $B$ regions; $B(\bm
r)=1$ when $\bm r$ lies within the $B$ regions and $B(\bm 
r)=0$ when $\bm r$ lies within the $A$ regions. The characteristic
function is periodic $B(\mathbf{r})=B(\mathbf{r}+\mathbf{R})$, with
$\{\mathbf{R}\}$ the Bravais lattice of the metamaterial.

According to Eq. (6) of Ref. \onlinecite{mochanOE10} the inverse macroscopic
longitudinal dielectric response of the system, defined through
$
  \mathbf{E}_{ML} = \boldsymbol{\epsilon}_{ML}^{-1}\cdot \mathbf{D}_{ML}
$,  is given by
\begin{equation}\label{epsinv}
    \boldsymbol{\epsilon}^{-1}_{ML}
= \hat{\mathbf{q}}\eta^{-1}_{\mathbf0\mathbf0}\hat{\mathbf{q}} =
\hat{\mathbf{q}}\xi\hat{\mathbf{q}}, 
\end{equation} 
where
\begin{equation}\label{xivseta}
  \xi\equiv\eta^{-1}_{\mathbf{00}}
\end{equation}
is the $\mathbf{00}$ component of the
inverse of the matrix
\begin{equation}\label{eta}
\eta_{\mathbf{G}\mathbf{G}'}\equiv \hat{\mathbf{G}} \cdot( \epsilon_{\mathbf{G}\mathbf{G}'} \hat{\mathbf{G}'}).
\end{equation}
Here, $\mathbf E_{ML}$ and $\mathbf D_{ML}$ are the longitudinal
projections of the macroscopic electric and displacement fields with
wavevector $\mathbf q$,
$\epsilon_{\mathbf{GG}}$ is the Fourier transform of the microscopic dielectric
response (Eq. \eqref{er}) with wave-vector $\mathbf G-\mathbf G'$, and
$\{\mathbf{G}\}$ is the reciprocal lattice of the system, so that
the microscopic constitutive equation
may be written in reciprocal space as 
\begin{equation}\label{DG}
  \mathbf{D}_{\mathbf{G}}(\mathbf{q}) = \sum_{\mathbf{G}'}\epsilon_{\mathbf{G}\mathbf{G}'} \mathbf{E}_{\mathbf{G}'}(\mathbf{q}),
\end{equation}
where $\mathbf{D}_{\mathbf{G}}(\mathbf{q})$ and
$\mathbf{E}_{\mathbf{G}}(\mathbf{q})$ are the 
Fourier coefficients of the fields with wavevectors
$\mathbf{q}+\mathbf{G}$. The wavevector $\mathbf{q}$ of the
macroscopic field may be interpreted as the conserved Bloch's vector. 
For succinctness, and in accordance to the long wavelength
approximation, we have  denoted the unit vectors
$(\mathbf{q}+\mathbf{G})/|\mathbf{q}+\mathbf{G}|$  simply by
$\hat{\mathbf{G}}$, and in particular, 
$\hat{\mathbf{0}}\equiv\mathbf{q}/q=\hat{\mathbf{q}}$. 

We should emphasize that $\xi$ and
$\boldsymbol\epsilon^{-1}_{ML}$ in
Eq. (\ref{epsinv})  
depend in general on the direction
$\hat{\mathbf{q}}$ of $\mathbf{q}$. Nevertheless,
\begin{equation}\label{epsilonM}
  \boldsymbol \epsilon_{ML} = \hat{\mathbf q} \xi^{-1} \hat{\mathbf q} = 
\hat{\mathbf q} \hat{\mathbf q} \cdot \boldsymbol \epsilon_M \cdot
\hat{\mathbf q} \hat{\mathbf q}. 
\end{equation}
is simply the longitudinal projection of
the macroscopic dielectric tensor
$\bm\epsilon_M$, and $\bm\epsilon_M$ is independent of the direction
$\hat{\bm q}$ in the long wavelength limit $\bm q\to0$. Therefore, 
calculating $\xi(\hat{\bm{q}})$, for 
several propagation directions $\hat{\bm{q}}$, 
we may
obtain \emph{all} the components of the long-wavelength
dielectric tensor 
$\bm\epsilon_M(0)$.\cite{nota2} 
For example, setting $\bm q$ along $\hat{\bm x}$,
Eq. \eqref{epsilonM} allows us to identify $\xi^{-1}(\hat{\bm
  x})=\epsilon_{M}^{xx}$. Similarly, 
setting $\bm 
  q$ along $\hat{\bm y}$ we obtain $\xi^{-1}(\hat{\bm
  y})=\epsilon_{M}^{yy}$, and setting 
$\bm q$ along $\hat{\bm x}+\hat{\bm y}$ we obtain
$\xi^{-1}((\hat{\bm x}+\hat{\bm
    y})/\sqrt2)=(\epsilon_{M}^{xx}+2\epsilon_{M}^{xy}+\epsilon_{M}^{yy})/2$
  from which we finally obtain $\epsilon_{M}^{xy}
  (=\epsilon_{M}^{yx})$ and thus, the 
full {\em transverse} dielectric tensor for waves propagating along
the $Z$ axis.    

We can calculate $\xi$, appearing in Eq. \eqref{epsinv}, very efficiently using
Haydock's recursion method, as shown in
Refs. \onlinecite{mochanOE10} and \onlinecite{cortesPSS10}, 
\begin{equation}\label{epsmh}
\xi=\frac{u}{\epsilon_a}
\frac{1}
{u-a_0-
\frac{b_1^2}
{u-a_1
-\frac{b_2^2}
{u-a_2
-\frac{b_3^2}
{\ddots}
}
}
}
,
\end{equation}
where $a_n$ and $b_n$ are Haydock's coefficients, defined through
\begin{equation}\label{keth}
|{\tilde{n}}\rangle=\hat{\cal H}|{n-1}\rangle=b_{n-1}|{n-2}\rangle + a_{n-1}|{n-1}\rangle + b_{n}|{n}\rangle
.
\end{equation}
From Eqs. (\ref{xivseta}) and (\ref{eta}), $\xi$ plays the role of a
Green's function projected onto the 
macroscopic {\em state} $|0\rangle$ corresponding to a longitudinal
plane wave with 
wave vector $\bm q$, the {\em spectral variable} $u(\omega)\equiv
(1-\epsilon_b(\omega)/\epsilon_a(\omega))^{-1}$ plays the role of a
(complex) {\em   energy} and
\begin{equation}\label{fll}
{\cal H}_{\bm{G}\bm{G'}}\equiv B_{\bm{G}\bm{G'}}^{LL}=\hat{\bm{G}}\cdot(B_{\bm{G}\bm{G'}}\hat{\bm{G'}})
\end{equation}
plays the role of a Hamiltonian, with
\begin{equation}\label{BGG}
B_{\bm{G}\bm{G'}}\equiv B_{\bm G-\bm G'}= \frac{1}{\Omega}\int
d^3r\,B(\bm r)e^{-i(\bm{G}-\bm{G'})\cdot\bm{r}}= \frac{1}{\Omega}\int_v
d^3r\,e^{-i(\bm{G}-\bm{G'})\cdot\bm{r}}
\end{equation}
the Fourier transform of the characteristic function
describing  the geometry of inclusions which occupy the volume $v$
within a unit cell of volume $\Omega$.
The recursion \eqref{keth} starts from the macroscopic state  $|0\rangle$ 
and we impose
the orthonormality condition $\langle n|m\rangle=\delta_{nm}$, where
$\delta_{nm}$ is Kronecker's delta function, to obtain the coefficients
$a_{n-1}=\langle n-1|\tilde{n}\rangle=\langle n-1|\hat{\cal
  H}|{n-1}\rangle$ and $b_{n}^2=\langle {\tilde n} |{\tilde n}\rangle
- a_{n-1}^2-b_{n-1}^2$. After calculating the macroscopic response
from Eq. \eqref{epsilonM} with different directions $\hat{\bm q}$,
optical 
properties such as
reflectance and absortance may be calculated using standard
formulae.\cite{barreraPRB07,BornWolf(1999)} 
Further details of this
calculation in the 2D and 3D case may be found in
Refs. 
\onlinecite{mochanOE10}
and
\onlinecite{cortesPSS10}, respectively.

For a generic system, all the Cartesian components of its macroscopic
dielectric function $\epsilon^{ij}_M$ might be
non-null, although some  of the off-diagonal components   
$\epsilon^{ij}_M$ with $i\neq j$ might be zero due to the symmetries of
the system.
For the the 2D metamaterials with translational symmetry along the
$Z$ axis that we study here, and for propagation along $Z$,
$\epsilon^{ij}_M$ might be described by a complex $2\times2$ tensor 
\begin{equation}\label{epsi}
\boldsymbol{\epsilon}_M
=\left(\begin{array}{cc}
\epsilon_M^{xx}  & \epsilon_M^{xy}   \\
\epsilon_M^{yx}   & \epsilon_M^{yy}   \\
\end{array}
\right)
.
\end{equation}  
Here $X$ and $Y$ are the Cartesian directions in a coordinate system 
fixed to the unit cell (see Fig.~\ref{cross}). In general
$\epsilon_M^{xx}\neq \epsilon_M^{yy}$ 
and $\epsilon_M^{xy}=\epsilon_M^{yx}\neq 0$.    
Thus, 
it is convenient to rotate the $XY$ Cartesian
system to the so called principal axes $X'Y'$ of the system in which  
$\epsilon^{i'j'}_M$ becomes diagonal.
We note that the direction of the principal axes depend in general on
the composition of the metamaterial and on the frequency, and are not
completely determined by its geometry. Furthermore, since $\boldsymbol{\epsilon}_M$ is
in general complex due to the presence of dissipation,
the vectors that define the principal axes are also
complex. This means that their real parts could point in
directions that differ from those of their imaginary parts, so that
there are no {\em real} directions in space along which the fields of the
corresponding eigenmodes oscillate. The polarizations of the
eigenmodes are thus elliptical in general. 

The eigenvalues $\lambda_\mu$ and eigenvectors $\mathbf{V}_\mu$ ($\mu=1,2$) of the
$2\times 2$  
symmetric (though complex) tensor 
$\boldsymbol{\epsilon}_M$, 
are obtained straightforwardly.\cite{arfken}
Notice that as the 
macroscopic response is not in general a self-adjunct matrix, its
eigenvectors are not 
orthogonal according to the Hermitian product.  Nevertheless, in the
nonretarded limit, the dielectric function is symmetric, so that its
eigenvectors are orthogonal according to the Euclidean
product. However, for convenience we normalize the eigenvectors
using the Hermitian norm $\mathbf{V}_\mu^*\cdot\mathbf{V}_\mu=1$
(otherwise, we would  
be unable to normalize the eigenvectors corresponding to circular
polarization). The eigenvalues $\lambda_\mu$ correspond to the principal values of
the macroscopic dielectric tensor $\boldsymbol{\epsilon}_M$, 
so the corresponding principal values of the complex index of refraction are
\begin{equation}\label{nmu}
n_\mu=\sqrt{\lambda_\mu}.  
\end{equation}

\subsection{Elliptical Polarization}\label{ep}

To describe the polarization ellipse corresponding to the eigenvectors
$\mathbf{V}_\mu$, 
or more generally, to describe the polarization of an arbitrary
monochromatic field
$\boldsymbol{\mathcal{\cal E}}(t)=\mathrm{Re}\left(\mathbf{E}_0 e^{-i\omega t}\right)$ with
frequency $\omega$, we separate the
complex amplitude $\mathbf{E}_0$ into real and  imaginary parts
\begin{equation}\label{w.1}
\mathbf{E}_0
=
\mathbf{E}_0' 
+i
\mathbf{E}_0''
,
\end{equation} 
to write 
\begin{align}\label{w.2}
\boldsymbol{\mathcal{\cal E}}(t)&=\mathbf{E}_0' \cos(\omega t) + \mathbf{E}_0''\sin(\omega t).
\end{align}
We interpret this real transverse vector equation as a $2\times2$ system of
equations which we solve for 
$\sin\omega t$ and $\cos\omega t$. Then, we write the
trigonometric identity $\cos^2(\omega t)+\sin^2(\omega t)=1$ as the
real quadratic form
\begin{equation}\label{w.6}
\boldsymbol{\mathcal{\cal E}}^T(t)\cdot \boldsymbol{\mathcal{\cal M}}\cdot \boldsymbol{\mathcal{\cal E}}(t)=1 
,
\end{equation}
where we interpret
$\boldsymbol{\mathcal{\cal E}}(t)$ as a column vector, $\boldsymbol{\mathcal{\cal E}}^T(t)$ as its transpose, and
$\boldsymbol{\mathcal{\cal M}}$ as a matrix with components 
\begin{subequations}
\begin{eqnarray}\label{w.8}
	   {\cal M}^{xx}&=&|E_{0 y}|^2/D,
           \\
	   {\cal M}^{xy}={\cal M}^{yx}&=&-(E'_{0 x} E'_{0 y} +E''_{0 x}
           E''_{0 y})/D,
           \\
           {\cal M}^{yy}&=&|E_{0x}|^2/D,
\end{eqnarray}
\end{subequations}
with
\begin{equation}\label{w.9}
D=(E'_{0 x} E''_{0 y} -E'_{0 y} E''_{0 x})^2
.
\end{equation}
The quadratic equation \eqref{w.6} describes the polarization ellipse,
which may be further characterized by diagonalizing $\boldsymbol{\mathcal{\cal M}}$,
i.e., solving $\boldsymbol{\mathcal{\cal M}}\cdot\boldsymbol{\mathcal{\cal V}}_{\pm}
=\Lambda_{\pm}\boldsymbol{\mathcal{\cal V}}_{\pm}$, to obtain the real positive eigenvalues
$\Lambda_{\pm}$ (choosing $\Lambda_+\geq\Lambda_-$)  
and their corresponding eigenvectors $\boldsymbol{\mathcal{\cal V}}_{\pm}$.
Writing $\boldsymbol{\mathcal{\cal E}}(t)=\boldsymbol{\mathcal{\cal E}}_{+}(t)\boldsymbol{\mathcal{\cal V}}_{+} +
\boldsymbol{\mathcal{\cal E}}_{-}(t)\boldsymbol{\mathcal{\cal V}}_{-}$ we obtain
\begin{equation}\label{w.51}
\boldsymbol{\mathcal{\cal E}}^T(t)\cdot \boldsymbol{\mathcal{\cal M}}\cdot \boldsymbol{\mathcal{\cal E}}(t)=
\Lambda_{-} {\cal E}_{-}^2(t) + \Lambda_{+} {\cal E}_{+}^2(t) = 1
,
\end{equation}
which is the equation of an ellipse with major $a$ and minor
$b$ semi-axes given by
\begin{subequations}
\begin{eqnarray}\label{w.69}
a&=&\frac{1}{\sqrt{\Lambda_-}}
\\
b&=&\frac{1}{\sqrt{\Lambda_+}}
.
\end{eqnarray}
\end{subequations}
The angles $\alpha$ and $\beta$ formed by the semi-axes $a$
and $b$ with respect to the $X$ axis are given by
\begin{subequations}
\begin{eqnarray}\label{w.70}
\tan\alpha&=&\frac{\Lambda_{-} -
  {\cal M}^{xx}}{{\cal M}^{xy}},
\\
\tan\beta&=&\frac{\Lambda_{+} - {\cal M}^{xx}}{{\cal M}^{xy}}.
\end{eqnarray}
\end{subequations}
As expected, the minor and major semiaxes are mutually orthogonal. 

The sense along which the field goes around the polarization ellipse
is determined from  the helicity $h=\mathrm{sgn}[\hat{\mathbf k}\cdot
  \mathrm{Re}(\mathbf{E}_0) \times\mathrm{Im}(\mathbf{E}_0)]$, with
$\hat{\mathbf k}$ a unit vector along the propagation direction. For
waves moving along the positive $Z$ axis,
\begin{equation}\label{w.89}
h=\mathrm{sgn}(E'_{0x} E''_{0 y} -E''_{0 x} E'_{0 y}),
\end{equation}
where $h=+1$ corresponds to right handed polarization $\circlearrowleft$ and
and $h=-1$ to left handed polarization $\circlearrowright$.
The degree of linearity or circularity of the polarization can be
characterized through the so
called {\it third flattening}, $\eta$ of the
polarization ellipses, defined through
\begin{align}\label{3rd}
\eta
=\frac{a-b}{a+b}
,
\end{align} 
The values of $\eta$ go from 1 for linear polarization to 0 for
circular polarization, with intermediate values corresponding to
elliptical polarization. 

The analysis above may be applied to each of the normal modes
of the macroscopic response of the metamaterial, as well
as to the incoming, transmitted or reflected waves.  

\subsection{Thin Layer}

Consider a monochromatic wave impinging normally from an isotropic
transparent medium $I$
into a film $F$ of width $d$ made of our metamaterial, from where it is
partially reflected back into medium $I$ and transmitted into a
transparent isotropic medium $T$. Media $I$ and $T$ are characterized
by their index of refraction $n_i$ and $n_t$, while the film is
characterized by the macroscopic tensor $\boldsymbol{\epsilon}_M$ discussed in
subsection \ref{mr}. We can project the electric field $\mathbf{E}_i$ of the
incident wave into the principal {\em directions} of the response of
the metamaterial 
\begin{equation}\label{dual3}
\mathbf{E}_i=\sum_\mu E_{i\mu}\mathbf{V}_\mu=\sum_\mu\mathbf{E}_i\cdot\tilde{\mathbf{V}}_\mu\mathbf{V}_\mu
,
\end{equation}
by introducing a dual basis
\begin{equation}\label{dual1}
\tilde{\mathbf{V}}_\mu = \frac {\mathbf{V}_{\overline\mu}^\perp}
      {\mathbf{V}_{\mu}\cdot\mathbf{V}_{\overline\mu}^\perp}
,
\end{equation}
where for any possibly complex vector $\mathbf{v}=(v^x, v^y)$ on the $XY$
plane we define  
$\mathbf{v}^\perp=(-v^y, v^x)$ as a perpendicular vector obtained by
rotating clockwise by $90^\circ$ on the plane, and where
$\overline\mu$ denotes the index complementary to index $\mu$, i.e.,
$\overline 1=2$ and $\overline 2=1$, so that
$\tilde{\mathbf{V}}_\mu\cdot\mathbf{V}_\mu=1$ and
$\tilde{\mathbf{V}}_{\overline\mu}\cdot\mathbf{V}_\mu=0$. Notice that 
the dual vectors are not necessarily ortho{\em normal}
according to neither the Euclidean nor the Hermitian scalar product.

Each of the
principal polarizations is conserved as the wave propagates along the
system, and for each of them the film has a well defined index of
refraction $n_\mu$ (Eq. \eqref{nmu}). Thus, we can obtain the optical
properties of the system using the standard formulae for the reflection and
transmission amplitudes of a thin film,\cite{BornWolf(1999)}  i.e.
\begin{subequations}\label{tl1} 
\begin{align}
r_\mu&=\frac{r_{i\mu}+r_{\mu t}e^{2ik_\mu
    d}}{1+r_{i\mu}r_{\mu t}e^{2ik_\mu d}}
\\ 
t_\mu&=\frac{t_{i\mu}t_{\mu t}e^{ik_\mu
    d}}{1+r_{i\mu}r_{\mu t}e^{2ik_\mu d}}
,
\end{align}   
\end{subequations}
where 
\begin{subequations}
\begin{align}\label{tl2}
r_{pq}&=\frac{n_p-n_q}{n_p+n_q}
\\
t_{pq}&=\frac{2n_p}{n_p+n_q}
,
\end{align}  
\end{subequations}
are the reflection and transmission coefficients corresponding to a
single interface separating 
medium $p$ from medium $q$ ($p,q=i,\mu,t$),
$d$ is the film thickness, and $k_\mu=(\omega/c)n_\mu$ is the wavenumber
within the film corresponding to the mode $\mu=1,2$.  

According to Eq. \eqref{dual3}, 
the reflected and transmitted electric fields are given by
\begin{subequations}\label{tl3}
\begin{align}
\mathbf{E}_r&=\sum_\mu r_\mu \mathbf{E}_i\cdot\tilde{\mathbf{V}}_\mu\mathbf{V}_\mu,
\\
\mathbf{E}_t&=\sum_\mu t_\mu \mathbf{E}_i\cdot\tilde{\mathbf{V}}_\mu\mathbf{V}_\mu
.
\end{align}
\end{subequations} 
The reflectance and transmittance are given by
\begin{subequations}\label{tl6}
\begin{align}
R&=\frac{|\mathbf{E}_r|^2}{|\mathbf{E}_i|^2},
\\
T&=\frac{n_t}{n_i}\frac{|\mathbf{E}_t|^2}{|\mathbf{E}_i|^2},
\end{align}
\end{subequations}
and the polarization properties of the incident, reflected and
transmitted field may be found through the analysis of subsection \ref{ep}.

The optical properties obtained above depend implicitly on the
frequency of the incident field through $n_\mu$, which inherits its
frequency dependence from the response $\epsilon_a(\omega)$ and $\epsilon_b(\omega)$
of the components $A$ and $B$ of the metamaterial.

\subsection{Polarization}

We can write Eqs. \eqref{tl3} as a matrix equation
\begin{equation}\label{Jalpha}
  \mathbf{E}_\alpha=\mathbf{J}_\alpha\cdot\mathbf{E}_i,\quad(\alpha=t,r),
\end{equation}
where
\begin{equation}\label{JalphaA}
\mathbf{J}_\alpha
=
\left(\begin{array}{cc}
V_{x1} & V_{x2}  \\
 V_{y1}  & V_{y2}   \\
\end{array}
\right) 
\left(\begin{array}{cc}
\alpha_1 & 0  \\
0 & \alpha_2  \\
\end{array}
\right) 
\left(\begin{array}{cc}
\tilde V_{1x} & \tilde V_{1y}  \\
 \tilde V_{2x}  & \tilde V_{2y}   \\
\end{array}
\right) 
\end{equation}
denote the Jones matrices\cite{Brosseau(1988)} for
reflection ($\alpha=r$) or transmission ($\alpha=t$), and where
$r_\mu$ and $t_\mu$ ($\mu=1,2$) are taken from Eqs. \eqref{tl1}. The Jones
matrices \eqref{JalphaA} allow us to calculate the polarization of the
reflected and transmitted light from the polarization of the incoming
wave, assumed to be in a pure polarized state. When the incoming wave
is not in a pure state but has an unpolarized component, we can
describe its polarization state 
in terms of the Stokes vectors $\mathbf{S}_\alpha$ for the reflected
($\alpha=r$) and the transmitted ($\alpha=t$) wave, which are related
to the incoming polarization state $\mathbf{S}_i$ through
\begin{equation}\label{eq.6}
\mathbf{S}_\alpha=\mathbf{M}_\alpha\cdot\mathbf{S}_i,\quad(\alpha=r,t).
\end{equation}
in terms of the Mueller matrices $\mathbf{M}_\alpha$, with
components\cite{Brosseau(1988)} 
\begin{align}\label{eq.4}
m^\alpha_{ij} = \frac{1}{2} \mathrm{Tr}
\left(\mathbf{J}_\alpha\sigma_i\mathbf{J}^\dagger_\alpha\sigma_j\right),
 \quad(\alpha=t,r),\quad(i,j=0,1,2,3),
\end{align}
with $\sigma_i$ the Pauli matrices plus the identity, i.e.
\begin{align}\label{qe.5}
\sigma_0=
\left(\begin{array}{cc}
1 & 0  \\
 0  & 1   \\
\end{array}
\right) 
\quad 
\sigma_1=
\left(\begin{array}{cc}
1 & 0  \\
 0  & -1   \\
\end{array}
\right) 
\quad 
\sigma_2=
\left(\begin{array}{cc}
0 & 1  \\
 1  & 0   \\
\end{array}
\right) 
\quad 
\sigma_3=
\left(\begin{array}{cc}
0 & -i  \\
 i  & 0   \\
\end{array}
\right) 
,
\end{align}
and where  $J_\alpha^\dagger$ denotes the adjunct
of the matrix $J_\alpha$. 
For example, for unpolarized (natural) incoming light, the input
Stokes vector would be
\begin{equation}\label{eq.90}
\mathbf{S}_i=(1, 0,0,0)^T,
\end{equation}
so the output Stokes vector would be
\begin{equation}\label{eq.90.2}
  \mathbf{S}_\alpha=
  (m^\alpha_{00}, m^\alpha_{10}, m^\alpha_{20}, m^\alpha_{30})^T, \quad(\alpha=r,t).
\end{equation}
Here, the superscript $T$ denotes transpose.

The degree of polarization of the outgoing waves can be described by
\begin{equation}\label{eq.89}
P_\alpha=\frac{\left(S^2_{\alpha 1}+S^2_{\alpha 2}+S^2_{\alpha 3}\right)^{1/2}}{S_{\alpha 0}}
\leq 1;
\end{equation}
$P_\alpha=1$ corresponds to fully polarized and $P_\alpha=0$
corresponds to unpolarized light. The kind of full or partial
polarization may be read from the Stokes coefficients: $S_{\alpha
  1}>0$ corresponds to (partial) linear horizontal polarization while 
$S_{\alpha 1}<0$ corresponds to vertical polarization, $S_{\alpha
  2}>0$ to polarization along $45^\circ$ from
the $x$ towards the $y$ axis, while $S_{\alpha 2}<0$ corresponds to  $135^\circ$,
$S_{\alpha 3}>0$ corresponds to circular right-handed polarization, and
$S_{\alpha 3}<0$ to circular left-handed polarization.

\section{Results}\label{res}

We first consider a metamaterial made out of a thin conducting film of
width $d$ 
deposited on an isotropic dielectric substrate with index of
refraction $n_t$ and with a square array of
holes ($\epsilon_b=1$) in the form of crosses, as those shown in
Fig.~\ref{cross}.
The 
geometry of the system is then characterized by the lengths $L_x$ and
$L_y$ of the beams of the cross along the $x$ and $y$ directions,
their widths $W_x$ and $W_y$ and the angle $\phi$ between one beam of
the cross and the $x$ axis. We can tune these parameters to design
systems with desired optical properties.

\begin{figure}
  \includegraphics{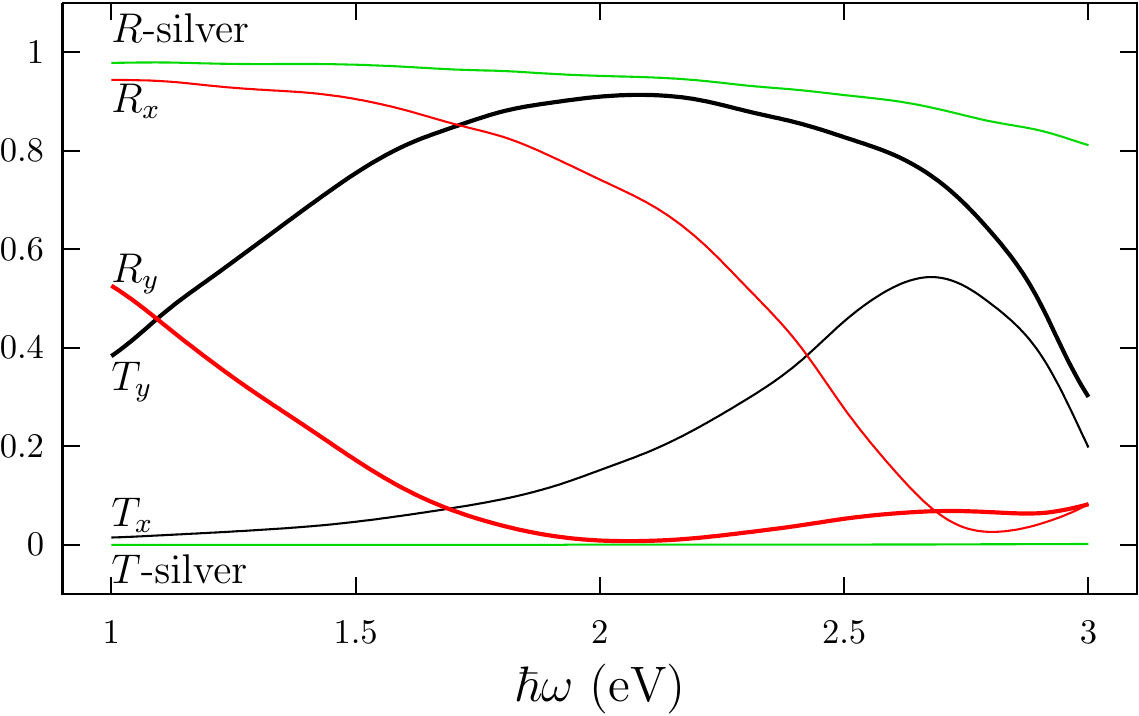}
  \caption{\label{linpol}Normal incidence reflectance $R_\mu$ and
    transmittance $T_\mu$ ($\mu=x,y$) of a
    $d=100$~nm Ag film on glass with a subwavelength square array of cross
    shaped holes with $L_x=0.965a$, $L_y=0.848a$, $W_x=0.299a$, and
    $W_y=0.374a$, with $a$ the lattice parameter,  for an 
    incident wave linearly polarized along $X$ and $Y$ directions as a
    function of photon energy. Also shown are the reflectance and
    transmittance for a uniform Ag film of width $d=50$~nm and thus,
    with the same amount of metal.}
\end{figure}

In Fig. \ref{linpol} we show the transmittance and reflectance 
$T$ and $R$ of a silver film  ($\epsilon_a=\epsilon_\mathrm{Ag}$) of width
$d=100$~nm deposited over glass ($n_t=1.4$) with a square array of cross-shaped
holes aligned with the lattice 
axes $\phi=0$ with geometry characterized by $L_x=0.965a$,  $L_y=0.848a$,
$W_x=0.299a$, and $W_y=0.374a$, where $a$ is the lattice parameter.
We show
results for normally 
incident linearly polarized incoming light,
\begin{align}\label{dual4}
\mathbf{E}_i=E_{oi}(\cos\theta,\sin\theta)
,
\end{align}
with $\theta$ the angle of polarization with respect to the  $X$ direction.
We display results for polarization along the $X$-axis ($\theta=0$) and
along the $Y$-axis ($\theta=90^\circ$). We performed this and the
following calculations using the recursive methods developed in
Ref. \onlinecite{mochanOE10} within a lattice of $601\times601$ pixels
and using between 200 and 400 Haydock's coefficients. The programs
were developed using the Perl Data Language
(PDL).\cite{glazebrook(1997),pdl2010} 

In this system,  $X$ and
$Y$ are symmetry directions, and therefore they coincide
with the principal directions of the macroscopic response. Thus, for
input polarization angles $\theta=0$, $90^\circ$ the
outgoing polarizations coincide with the incoming polarizations. Notice
that there is a sizable energy range around 2eV for
which $R_y$ is below 1\% and $T_y$ is above 80\%, while, rotating the
incoming polarization by $90^\circ$, $R_x$ becomes much larger than 
$T_x$. Thus, this system displays a very large linear dichroism
both under transmission and under reflection. The reason for this
behavior is the extraordinary transmission  present in
conducting films whenever the conducting paths are almost
chocked, and is due to the matching between the vacuum surface impedance and
the surface impedance of the film as it transits from being
conductor-like at low frequencies and dielectric-like at high
frequencies.\cite{Mendoza(2012),mendozaPRB12a} Notice that for this system,
$P_x=0.035a$ and $P_y=0.152a$, so it has relatively wide conducting
paths along the $x$ 
direction, but very narrow passages along the $y$ direction. Therefore, the
film displays extraordinary transmission for
$y$ polarization but is opaque and thus has a large reflectance for
$x$ polarization. Thus, its extreme dichroism. For comparison
purposes, in the same figure we have plotted the transmittance of a
much thinner flat homogeneous Ag film, of width $d=50$~nm chosen so both
films contain 
the same amount of silver. We remark that the transmittance $T_y$ for $y$
polarization is about three orders of magnitude larger than that of the
homogeneous film, although the latter is narrower. 

\begin{figure}
\includegraphics{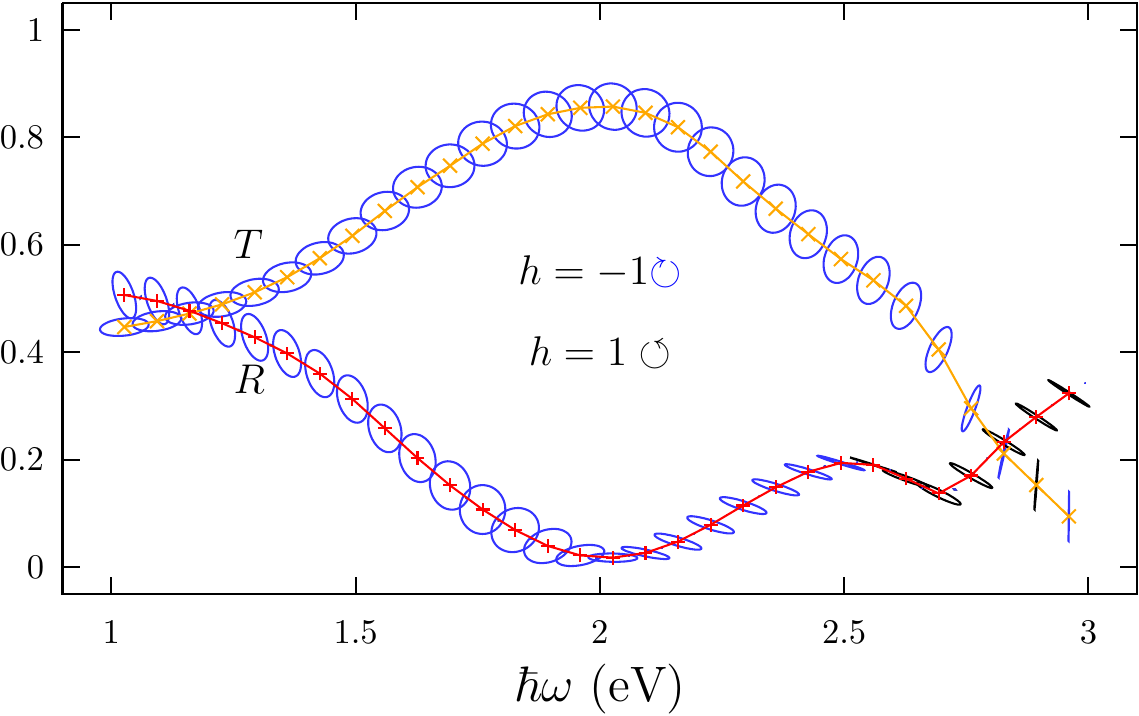}
\caption{Normal incidence reflectance $R$ and transmittance $T$ for
  the same system as in Fig. \ref{linpol} but with $L_x=0.963a$,
  $L_y=a$, $W_x=0.249a$ and $W_y=0.324a$ as a function of the photon
  energy. The incoming wave is linearly polarized with
  $\theta=45^\circ$. We indicate the polarization ellipses of
  the outgoing waves and we color code their helicities.
}
\label{fig0}
\end{figure}

In Fig.~\ref{fig0} we show the normal incidence transmittance and reflectance 
$T$ and $R$ of a system with $L_x=0.963a$, $L_y=a$, $W_x=0.249a$ and
$W_y=0.324a$.
In contrast 
to the previous case, conducting paths along $x$ are completely closed
instead of being wide open as in Fig. \ref{linpol}. The conducting paths
along $y$ are almost blocked, as $P_x=0.07a$. In this case, the
system behaves as an anisotropic insulator unless the frequency is low
enough that the response along $y$ becomes conductor-like. The
frequency of this dielectric-conductor change of behavior may be tuned
by changing the width $P_x$ of the narrow channels. 
The reflectance and transmittance and
the polarization of the outgoing light depend on the input
polarization. The results displayed in Fig. ~\ref{fig0} correspond to
incoming light that is linearly polarized at an angle
$\theta=45^\circ$ with respect to the $x$-axis. In the same figure we
indicated the shape of 
the polarization ellipses corresponding to the outgoing waves and we
color coded their helicity. Notice that for a very wide frequency range
we were able to obtain right handed almost circularly polarized
transmitted light with a transmittance  $T$ above $70\%$. Given the
geometrical symmetry of this system, as the incoming polarization
angle $\theta$ diminishes towards 0 or increases towards $90^\circ$ the
outgoing polarization becomes linear, and as $\theta$ approaches
$-45^\circ$ we obtain again circularly polarized light but with the
opposite helicity. Thus our system behaves as a
quarter-wave plate, though it is of subwavelength thickness and is
operational over a wide frequency range.

\begin{figure}
  \includegraphics{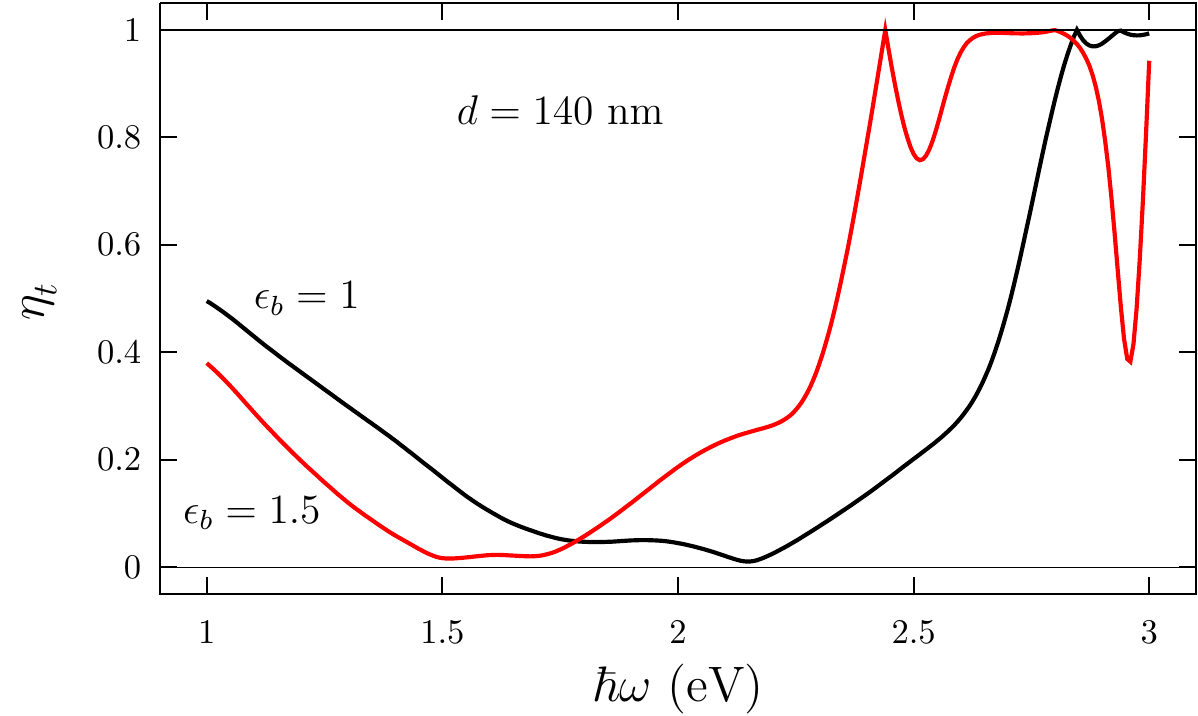}
  \caption{Third flattening $\eta_t$ for the transmitted
    fields corresponding to Fig. \ref{fig0} illuminated by a field that
    is linearly polarized at
    $\theta=45^\circ$ or $135^\circ$. The film has a width
    $d=140$~nm and its holes are filled with an insulator with
    dielectric constant $\epsilon_b$.
 \label{etas}}
\end{figure}

In Fig. \ref{etas} we show the third flattening $\eta^{\strut}_t$ that
characterizes the degree of circular polarization of the
transmitted fields corresponding to the same system as in
Fig. \ref{fig0}, for
incident light linearly polarized along $\theta=45^\circ$ or
$135^\circ$. The film is wider $d=140$~nm and the holes are filled with
a transparent material with dielectric constant $\epsilon_b$.
We notice the wide energy regions for which $\eta^{\strut}_T$ is close to zero
(say, $<0.1$) 
and the fact that these regions may be shifted by changing the
dielectric constant $\epsilon_b$ of the insulating material. 
We have verified that the results above remain valid qualitatively
under changes of the geometrical parameters, although the angle of the
incoming polarization to obtain circular polarization might have to be
adjusted away from $45^\circ$, $135^\circ$,  and the energy range for
which we obtain circular polarization may also change.

\begin{figure}
  \includegraphics{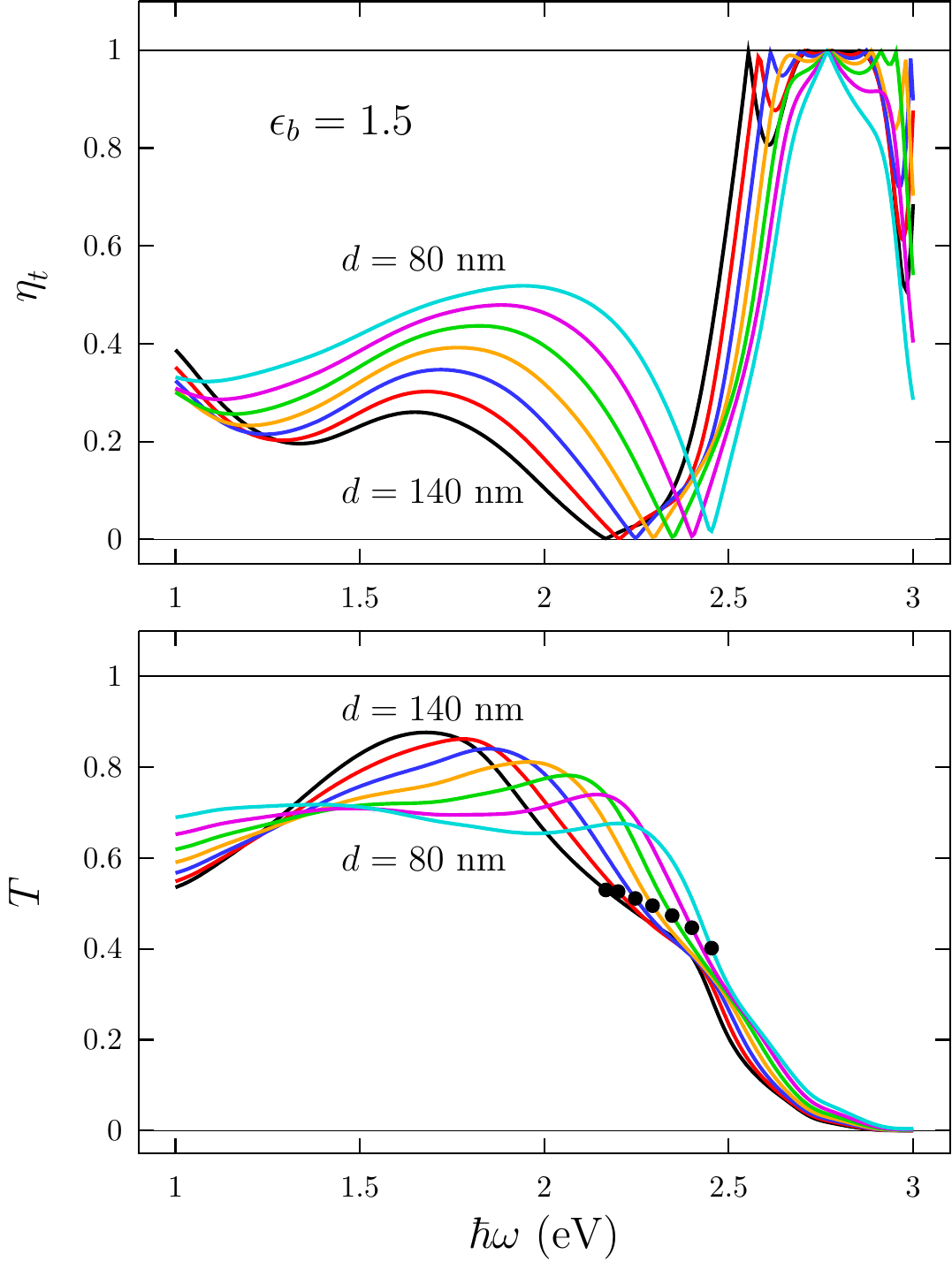}
  \caption{Third flattening $\eta^{\strut}_t$ (upper panel) for the transmitted
    light and transmittance $T$ (lower 
    panel) for the same film as in Fig. \ref{etas} with the holes
    filled by an insulator with dielectric constant $\epsilon_b=1.5$
    and for different film widths $d=80,90\ldots140$~nm, when illuminated by
    light linearly polarized along $\theta=35^\circ, 145^\circ$. The
    black dots in the lower panel correspond to the crossings
    $\eta^{\strut}_t=0$ in the upper panel.
    \label{etaT}}
\end{figure}
In Fig. \ref{etaT} we show the third flattening $\eta^{\strut}_t$ and the
transmittance $T$ for light transmitted through a film as in
Fig. \ref{etas} but of different widths $d=80,90,\ldots140$~nm when
illuminated by light linearly polarized along $\theta=35^\circ$ or
$145^\circ$. The holes in the Ag film are filled with an insulator
with dielectric constant $\epsilon_b=1.5$. Notice that fully circular
polarization ($\eta^{\strut}_t=0$) is achieved at a frequency that may be shifted
by $0.3$~eV by changing the width of the film. The black dots in the
lower panel show that the transmittance is appreciable ($T>0.3$) at
those frequencies for which the
transmitted field is completely circularly polarized. Similar
results hold for other dielectrics and other polarization angles and
it is possible to 
design the system to produce circularly polarized light at any visible
frequency. 

The systems analyzed above are symmetric under $x\to-x$ and $y\to-y$
reflections, and therefore, they display no circular dichroism and
they yield no circular polarization when
illuminated by natural, unpolarized light. Thus, to explore
metamaterials with circular dichroism and circular polarizers, we now
consider a system that has no in-plane reflection symmetry.
\begin{figure}
  \includegraphics[width=.7\textwidth]{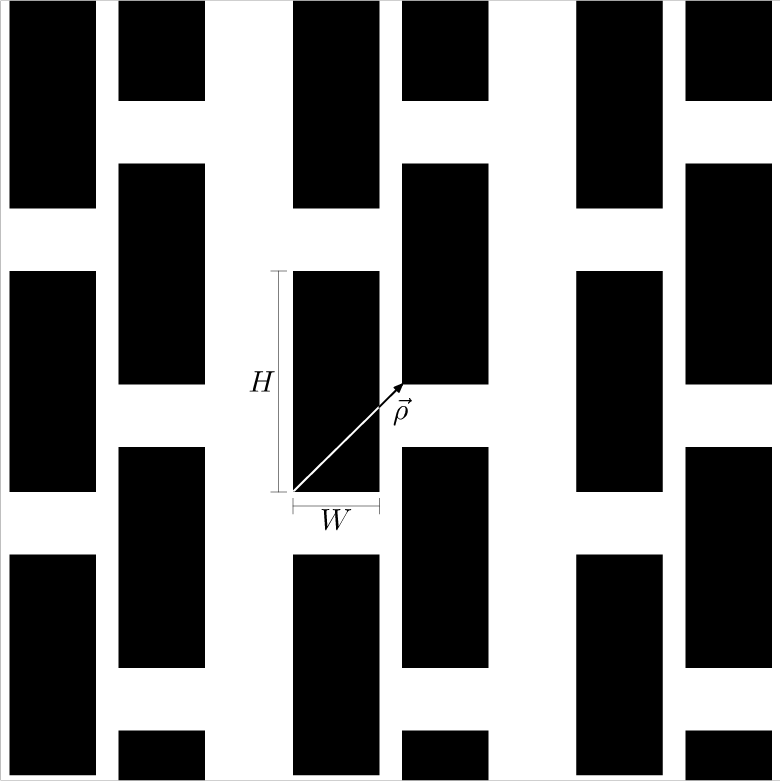}
  \caption{Ag film (white) deposited on a
glass substrate and from which a square lattice of pairs of holes in
the shape of rectangular prisms have been carved 
out and filled with a dielectric (black). The geometry is
characterized by the width $W$ and height $H$ of 
the prisms and the relative displacement $\boldsymbol \rho$ between pairs of
holes within the unit cell. For some values of these parameters the
dielectric filled holes might overlap each other.
\label{circularOptGeom}} 
\end{figure}

In Fig. \ref{circularOptGeom} we show
one such system, consisting of an Ag film deposited on a
glass substrate and from which a square lattice of pairs of 
holes have been carved out and filled with a dielectric. The holes
have the shape of prisms 
characterized by their width $W$ and height $H$, and relative
displacement $\boldsymbol \rho$ between pairs of holes within each
cell besides
the width $d$ of the film and the dielectric function 
$\epsilon_b$ of the inclusions. We optimized these parameters, as well
as the width $d$ of the film and the dielectric constant $\epsilon_b$
of the inclusions in order
to maximize the sought optical properties of the film. Notice that for
some values of the parameters, the dielectric filled holes might
overlap each other; our calculation procedure is able to cope with such
situations. 

\begin{figure}
  \includegraphics{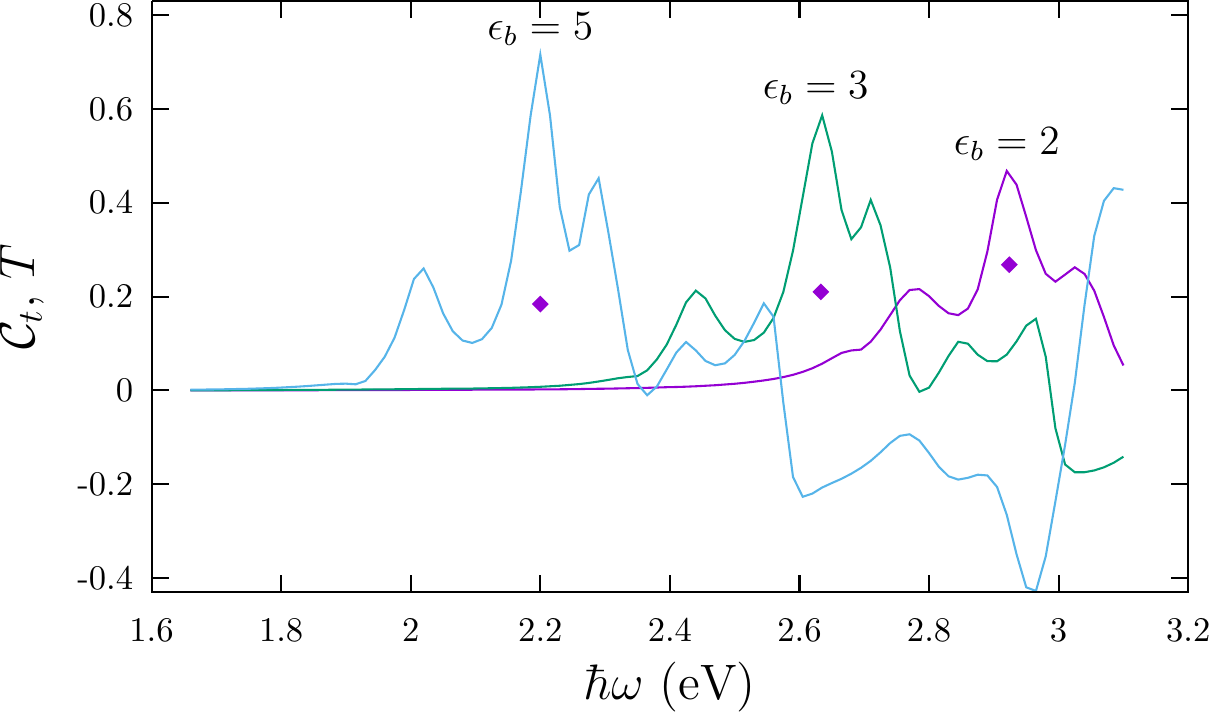}
  \caption{Degree of circular
    polarization $\mathcal C_t$ of light
    transmitted through a $d=100$~nm Ag film 
    with a geometry as shown in Fig. \ref{circularOptGeom} with
    parameters given in table \ref{tabla1} when illuminated by
    unpolarized light $S_i=(1,0,0,0)$. We display with solid circles
    the maxima of $\mathcal C_t$ and with solid diamonds the
    corresponding transmittance $T$.
\label{circularOptA}}
\end{figure}

In Fig. \ref{circularOptA} we show the degree of circular polarization
$\mathcal C_t\equiv S_{t3}/S_{t0}$, 
of light transmitted through a $d=100$~nm Ag and dielectric
metamaterial film on glass, as in
Fig. \ref{circularOptGeom}, when illuminated by unpolarized light,
corresponding to $S_i=(1,0,0,0)$. We show results for given values of the
dielectric constant $\epsilon_b=2,3,5$ and the geometrical parameters, given
in table \ref{tabla1}, were
obtained by finding the maximum circularity within the visible
range and optimizing it.
\begin{table}
  \begin{tabular}{ccccc}
    $\epsilon_b$&$W/a$&$H/a$&$\rho_x/a$&$\rho_y/a$\\
    \hline
    2&0.305&0.732&0.383&-0.304\\
    3&0.305&0.777&0.385&-0.379\\
    5&0.305&0.779&0.385&-0.379\\
  \end{tabular}
  \caption{Optimized geometrical parameters of the system shown in
    Fig. \ref{circularOptGeom} to produce the largest circular
    polarization  $\mathcal C_t$ within the visible range (see
    Fig. \ref{circularOptA}) 
    with a large enough transmittance $T$ for various values of the 
    dielectric constant $\epsilon_b=2,3,5$ for a film of thickness
    $d=100$~nm. over a glass substrate. The distances are expressed as
    fractions of the lattice parameter $a$.\label{tabla1}} 
\end{table}
Thus,  at each step during the optimization procedure we 
calculated the full spectra for each set of candidate parameters. To
guarantee a not-too-low transmittance, we actually maximized a mixture
of the degree of circular polarization 
$\mathcal C_t$, and the transmittance $T=S_{t0}$, to wit,
the product of two sigmoidal functions  of width 0.1 evaluated at
$\mathcal C_t$ and at $T$, and centered at $0.8$ and $0.2$, respectively. 
As illustrated in Fig. \ref{circularOptA}, the degree of circular
polarization may attain peaks with $\mathcal C_t=0.7$ or even higher,
with a corresponding transmittance larger than $\approx0.2$, and these peaks
may be tuned within the whole visible range by adequately choosing the
value of the dielectric constant $\epsilon_b$.
We remark that our computational scheme is fast enough to allow the
calculation of the full spectra {\em at each step} of the optimization
process. We performed the optimizations using the simplex method
offered by the MINUIT package 
developed at CERN \cite{James(1994)} and its PDL
\cite{glazebrook(1997),pdl2010} interface.
\cite{Jordan(2007)}

\begin{figure}
  \includegraphics{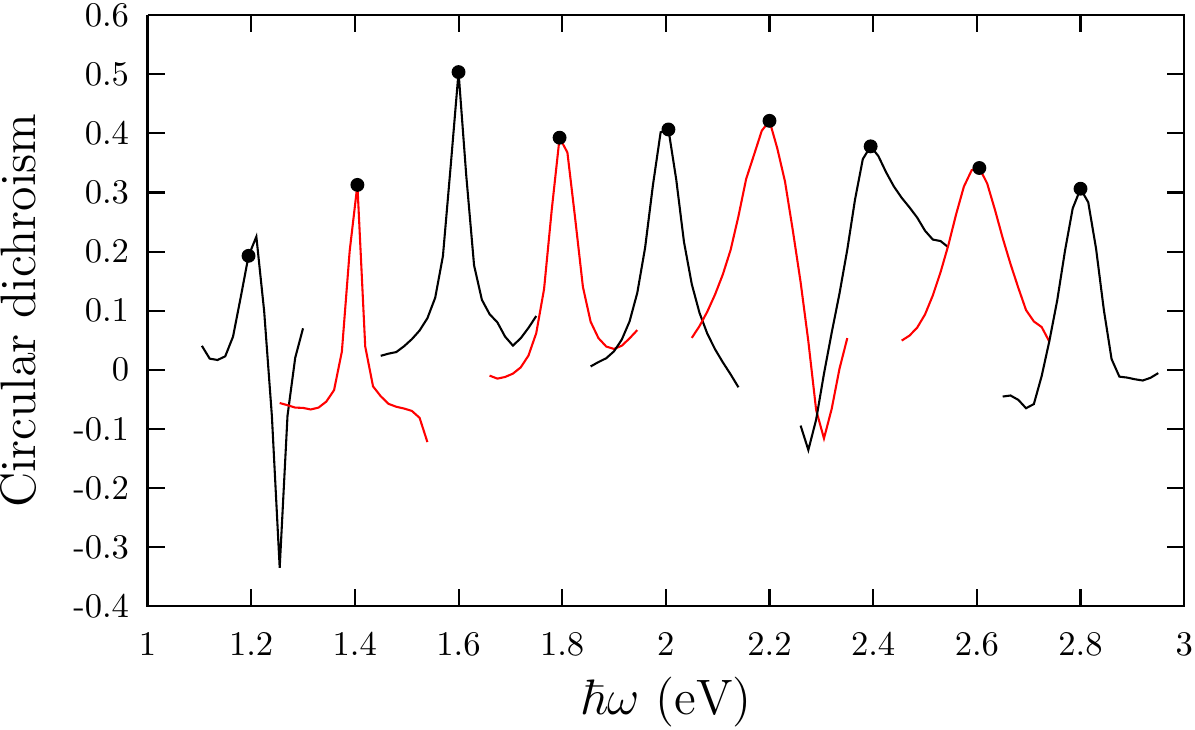}
  \caption{Circular dichroism $CD$ of a series of
    films with the geometry 
    described by Fig. \ref{circularOptGeom} and the parameters shown
    in table \ref{tablaDicroismo}. We only show part of the spectra
    around the energies $\hbar\omega_0$ for which the parameters were
    optimized, and we indicate with the solid dots the corresponding
    optimized values $CD(\omega_0)$. 
\label{dicroismoCircular}}
\end{figure}

Instead of searching for a maximum within a range of frequencies, we
can search for a maximum of any desired optical property at given
desired frequencies. To  illustrate this case, 
in Fig. \ref{dicroismoCircular} we show the circular dichroism
$CD=\mathcal A_L-\mathcal A_R$ of a
set of films with the geometrical parameters, as well as the
dielectric constants $\epsilon_b$ and the thickness $d$ of the film,
obtained by optimizing $CD$ at chosen frequencies
$\hbar\omega_0=1.2$~eV, $1.4$~eV\ldots $2.8$~eV and given in table
\ref{tablaDicroismo}. Here, $\mathcal A_L$
($\mathcal A_R$) is the absortance of the film corresponding to left-handed
(right-handed) circularly polarized incident light.  
\begin{table}
  \begin{tabular}{ccccccc}
    $\hbar\omega_0$ (eV)&$W/a$&$H/a$&$\rho_x/a$&$\rho_y/a$&$\epsilon_b$&$d$
    (nm)\\ 
    \hline
    1.2&0.491&0.815&0.499&-0.147&6.635&200\\
    1.4&0.490&0.793&0.461&-0.140&6.625&200\\
    1.6&0.415&0.652&0.434&0.191&6.939&171\\
    1.8&0.500&0.677&0.452&-0.214&4.893&198\\
    2.0&0.472&0.870&0.451&-0.213&4.842&166\\
    2.2&0.478&0.874&0.462&0.220&5.390&172\\
    2.4&0.485&0.844&0.483&0.282&5.470&172\\
    2.6&0.490&0.827&0.473&0.297&5.178&175\\
    2.8& 0.470&0.788&0.481&-0.302&5.465&177
  \end{tabular}
  \caption{Parameters (width $W$, height $H$, displacement
    $(\rho_x,\rho_y)$, dielectric constant $\epsilon_b$ and film
    thickness $d$) that optimize at the chosen frequency
    $\omega_0$ the circular dichroism of an Ag film
    with dielectric inclusions. The geometry and the parameters are as
    in Fig. \ref{circularOptGeom}. }
  \label{tablaDicroismo}  
\end{table}
Notice that with our simple geometry we obtained a circular dichroism
that peaks at our chosen frequencies which we tuned across the near
infrared and the visible region. Its maximum values are larger than
$0.2$ and as large as $0.5$, much larger than  those of naturally
occurring chiral materials, although our system is a thin film, its texture has
subwavelength characteristic distances, its geometry is not
chiral\cite{Kaschke(2015)} and the incoming light is normally
incident.\cite{Cao(2015)}  

\section{Conclusions}\label{conc}

We employed Haydock's recursive method within the long wavelength 
approximation to calculate the complex
frequency dependent macroscopic dielectric tensor 
$\epsilon^{ij}_M$
 of
metamaterials in terms of the  
dielectric functions of the host $\epsilon_a$ and the inclusions
$\epsilon_b$, and of 
the geometry of both the unit cell and the inclusions. 
The calculation requires modest
computing resources to obtain well converged results which can be applied
to metamaterials with dispersive and dissipative as well
as transparent components.  
The input to our calculations are
images of the unit cells which may be manipulated using image
processing software, thus allowing us to rapidly explore manifold
geometries in a design 
process to obtain a tailored optical response. 

We found that a simple system made up of a square array of
cross-shaped nanometric holes with slightly anisotropic geometrical
properties carved out of a thin supported silver film may display a very
strong linear dichroic response for both transmission and
reflection. Rotating the direction of polarization of the incoming
wave, the transmittance of the system could change from the very small
value expected of homogeneous Ag films to an extraordinary
transmittance that is about three orders of magnitude larger. 
Furthermore, the same system but with different
geometrical parameters behaves as a quarter wave plate, producing
circularly polarized output light for a linearly polarized input field
with an helicity that can be controlled by rotating the input
polarization direction, although the width of the film is much thinner
than the wavelength. Moreover, this behavior may be tuned over a wide
frequency range 
that covers the visible spectrum.

We also explored systems with no reflection symmetry within the
surface of the film and we were able to tune the parameters in order
to optimize different optical properties related to the circular
polarization of light. Thus, we obtained that a thin Ag film crossed by a
lattice of appropriately patterned insulating regions could behave as
a circular polarizer, yielding 
circularly polarized light when illuminated by unpolarized light, with
peak degrees of circular polarization above 0.7, and that a similar
system with different parameters yielded a film with extreme
circular dichroism as large as 0.5, much higher than that of naturally
occurring chiral materials. We remark that our system is a thin film,
its texture has subwavelength characteristic distances, its
geometry is not chiral and the incoming light is normally incident. 

These examples illustrate how standard optical elements may be replaced
by thin nanometric patterned films with the same or better
performance, which may thus be integrated into nano-photonic
devices. The design of these elements benefit greatly from the availability
of our very efficient computational procedure, which allowed us to
optimize the geometrical parameters of the system. 

\section*{Acknowledgments}
We acknowledge partial support from
DGAPA-UNAM through grant IN113016 (WLM) and from CONACyT 153930 (BSM). We are
grateful to J. Samuel P\'erez-Huerta and Guillermo Ortiz for useful
discussions.  

\bibliography{referencias}

\begin{thebibliography}{45}
\bibitem[{not()}]{nota2}
\bibinfo{note}{Since the long-wavelength response $\bm\epsilon_M$ is
  independent of the direction of $\bm q\to0$, one can use the results of the
  longitudinal calculations to solve optical (transverse) problems.}
\end{thebibliography}

\end{document}